# Generation of a VUV-to-visible Raman frequency comb in H₂-filled kagomé photonic crystal fiber

M. K. Mridha\*, D. Novoa, S. T. Bauerschmidt, A. Abdolvand and P. St.J. Russell

*Max Planck Institute for the Science of Light, Guenther-Scharowsky Str. 1, 91058 Erlangen, Germany*
*\*Corresponding author: manoj.mridha@mpl.mpg.de*

**We report the generation of a purely vibrational Raman comb, extending from the vacuum ultraviolet (184 nm) to the visible (478 nm), in hydrogen-filled kagomé-style photonic crystal fiber pumped at 266 nm. Stimulated Raman scattering and molecular modulation processes are enhanced by higher Raman gain in the ultraviolet. Owing to the pressure-tunable normal dispersion landscape of the "fiber + gas" system in the ultraviolet, higher-order anti-Stokes bands are generated preferentially in higher-order fiber modes. The results pave the way towards tunable fiber-based sources of deep- and vacuum ultraviolet light for applications in, e.g., spectroscopy and biomedicine.**

Fiber delivery and frequency conversion in the ultraviolet are topics of much current interest [1,2]. Step-index fibers with pure silica cores have proven unsuitable because their performance dramatically falls over time due to color-center-related optical damage, unless they are hydrogen loaded [3]. Recent alternatives such as fluoride fibers [4] partially resolve this problem, extending the usable wavelength range into the deep ultraviolet (DUV). Kagomé-style hollow-core photonic crystal fiber (kagomé-PCF) is excellent alternative, offering a very low light-glass overlap [5]. It has been used for stable long-term transmission of continuous-wave 280 nm DUV light [6]. Kagomé-PCFs also offer very broad spectral transmission windows, and when gas-filled provide an ideal system for nonlinear generation of light at many different wavelengths. For example, when pumped at near infrared wavelengths, broadband DUV and vacuum ultraviolet (VUV) light has been generated in gas-filled kagomé-PCF [7,8].

In this Letter, we discuss the first frequency-conversion scheme involving a H₂-filled kagomé-PCF pumped at 266 nm. Taking advantage of the high Raman gain of H₂ in the ultraviolet (about 3 times larger than at 532 nm [9]), we generate a purely vibrational Raman comb extending from the VUV (184 nm) to the visible (478 nm) via stimulated Raman scattering (SRS) and molecular modulation [10-13]. The short-wavelength bands extend further into the VUV than in previous experiments using H₂-filled kagomé-PCF with narrow-band pump lasers [14, 15]. Moreover, the narrow Raman gain bandwidth, together with a narrow-band pump, results in Raman sidebands with high spectral power density—a distinct advantage over broadband sources, when it is necessary to use filters (not always available in the DUV and VUV) to select spectral bands. The system demonstrated here is very compact compared to free-space arrangements [16] and its potential tunability using different Raman-active gases makes it ideal for applications such as high-resolution spectroscopy, metrology and biology.

In SRS a relatively strong pump pulse scatters off the molecules in the medium giving rise to a lower energy Stokes signal. The beat note of the pump and Stokes signals excites a Raman-active mode of the molecule (in our system, the fundamental vibrational mode of H₂ with $\Omega_R/2\pi \sim 125$ THz). This process is accompanied by the excitation of a coherence wave ($C_W$) of synchronous molecular oscillations in the gas, which in turn enhances the coherent amplification of the Stokes signal. Provided phase-matching is satisfied, these optical phonons can also trigger the efficient generation of anti-Stokes photons upshifted by $\Omega_R/2\pi$. Pressure-tunable dispersion in kagomé-PCF [17], augmented by the use of higher order modes [18], offers a route to perfectly phase-matched nonlinear frequency conversion in a collinear geometry. This is illustrated in Fig. 1, where pump photons (frequency $\omega_P$) propagating in a given fiber mode are scattered to Stokes photons (frequency $\omega_S$) in the same or a different mode, respectively generating intramodal or intermodal $C_W$'s [18]. These $C_W$'s, represented by colored arrows in Fig. 1, can then trigger the generation of further sidebands provided phase-matching and spatial overlap of the different transitions are met.

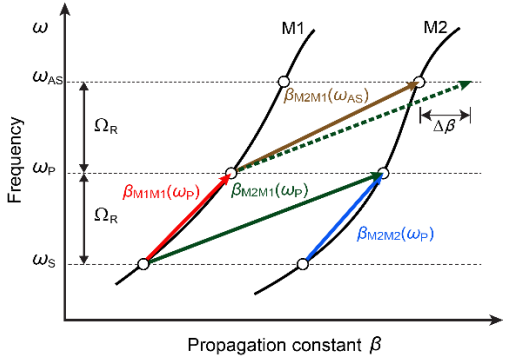

Fig. 1. Schematic representation of the dispersion curves of two hypothetical fiber core modes M1 and M2. The solid-red and solid-blue arrows represent intramodal $C_W$'s and the solid-green arrow represents an intermodal $C_W$, which can be used to up-convert photons from the pump in M1 to the AS in M2 (dashed-green arrow) provided the dephasing with the transition depicted by the solid-brown arrow is $\Delta\beta \sim 0$.

Each $C_W$ has frequency $\Omega_R/2\pi$ and wavevector $\beta_{12}(\omega) = \beta_1(\omega) - \beta_1(\omega - \Omega_R)$, where $\beta_1(\omega)$ and $\beta_2(\omega - \Omega_R)$ are the propagation constants of modes 1 and 2. As an example of the parametric nature of the anti-Stokes generation process, we see in Fig. 1 that the intermodal $C_W$ with wavevector $\beta_{M2M1}(\omega_P)$ (solid/dashed green arrow) may be used for efficient up-conversion of pump photons in mode M1 to anti-Stokes photons in mode M2 (solid-brown arrow), provided the dephasing $\Delta\beta = |\beta_{M2M1}(\omega_P) - \beta_{M2M1}(\omega_{AS})|$ is sufficiently small to make the dephasing length $2\pi/\Delta\beta$ longer than the fiber length.

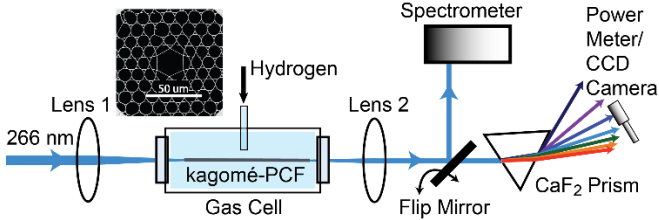

Fig. 2. Experimental set-up. The inset shows the scanning electron micrograph of the transverse structure of the fiber.

The experimental set-up is sketched in Fig. 2. We launched linearly-polarized, narrow-band DUV pulses (~3 ns duration, ~10 µJ pulse energy) at 266 nm into a 10-cm-long $H_2$-filled kagomé-PCF (core radius ~11.6 µm, see scanning electron micrograph in Fig. 2) with an ~80% coupling efficiency. At a pressure of 25 bar, 8 spectral sidebands from the 4th Stokes (S4) to the 4th anti-Stokes (AS4) were observed at the output of the fiber as shown in Fig. 3.

After dispersing the Raman comb with a $CaF_2$ prism, we measured the power of the comb lines with a calibrated power meter and imaged the different signals with a camera (see Fig. 3(a)). All the signals showed increasing higher-order mode (HOM) content with higher sideband order (both Stokes and anti-Stokes). The origin of this effect, and the key role it plays in the emission process, will be discussed below. To optimize the efficiency of the ultraviolet Raman comb, we monitored the output powers of 7 bands (S3-AS3 including the pump) while varying the pressure from 6 to 30 bar (see Fig. 4(a)). Interestingly, the power in each sideband peaks at a different pressure. This is summarized in Table 1 where we list the maximum conversion efficiencies obtained for the individual lines (defined as the ratio of the maximum power in each band to the pump power launched into the fiber). These pressure-dependent conversion efficiencies indicate the role of phase-matching in the efficient generation of fiber-based Raman combs.

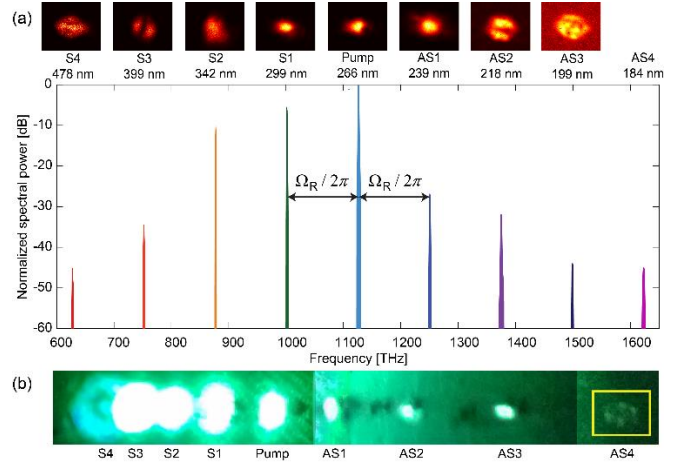

Fig. 3. (a) Measured spectrum of the Raman comb from S4 to AS4 at a pressure of 25 bar with 10 µJ pump energy. The peaks of the spectral lines were calibrated using a power meter and normalized to the peak of the pump spectrum. The spectral powers of the S4 and AS4 bands are arbitrarily set as they were below the sensitivity of the power meter used. Near-field optical images of the modal patterns at the fiber end face are shown above the plot (AS4 was too weak to be directly imaged). (b) Photograph of the spectrum in (a) dispersed at a $CaF_2$ prism and cast onto a fluorescent screen. The AS4 signal, highlighted by the yellow box in (b), exhibits a complex far-field profile.

In order to obtain deeper insight into the complex dynamics of comb generation in the ultraviolet, we numerically modeled the system using the following multimode-extended set of coupled Maxwell-Bloch equations [19,20]

$$\left(\frac{\partial}{\partial z} + \frac{1}{2}\alpha_{\sigma,l}\right)E_{\sigma,l} = -\kappa_{2,l}\frac{\omega_l}{\omega_{l-1}}i\sum_{\nu\xi\eta}^{M}\frac{S_{\sigma\nu\xi\eta}}{S_{\nu\xi}}Q_{\nu\xi}E_{\eta,l-1}q_{\eta,l-1}q_{\sigma,l}^* \quad (1)$$
$$-\kappa_{2,l+1}i\sum_{\nu\xi\eta}^{M}\frac{S_{\sigma\nu\xi\eta}}{S_{\nu\xi}}Q_{\nu\xi}^*E_{\eta,l+1}q_{\eta,l+1}q_{\sigma,l}^*$$

$$\frac{\partial}{\partial\tau}Q_{\nu\xi} = -Q_{\nu\xi}/T_2 - i(S_{\nu\xi}/4)\sum_l \kappa_{1,l}E_{\nu,l}E_{\xi,l-1}^*q_{\nu,l}q_{\xi,l-1}^*, \quad (2)$$

where $l$ denotes the sideband order with frequency $\omega_l = \omega_P + l\Omega_R$ and the summation is over all permutations of the modal set $M$. The electric field amplitude of mode $\sigma$ is $e_{\sigma,l}(z,\tau) = F_\sigma(x,y)E_{\sigma,l}(z,\tau)q_{\sigma,l}$, where $F_\sigma(x,y)$ is the normalized transverse spatial profile, $E_{\sigma,l}(z,\tau)$ is the slowly-varying envelope and $q_{\sigma,l} = \exp[-i\beta_\sigma(\omega_l)z]$ is the fast oscillating phase. $Q_{\nu\xi}$ represents the amplitude of the intramodal and intermodal $C_W$'s. The coupling constants $\kappa_{1,l}$ and $\kappa_{2,l}$ and the dephasing time $T_2$ of the Raman coherence are derived from experimental data [8]. The terms $\alpha_{\sigma,l}$ represent the fiber loss of the different signals. The overlap integrals $S_{\sigma\nu\xi\eta}$ and $S_{\nu\xi}$ are defined in [20]. We also assume that the majority of the molecules remain in the ground state.

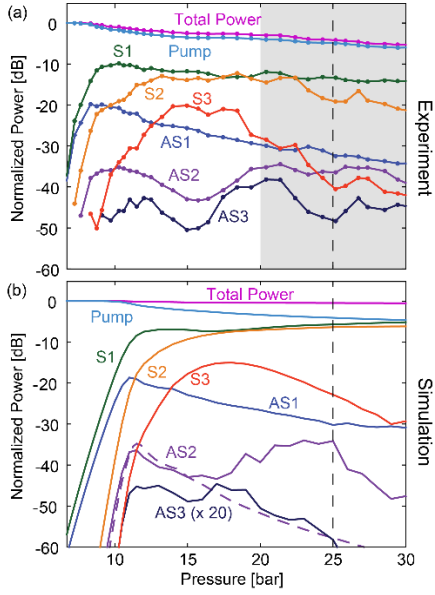

Fig. 4. (a) Measured and (b) numerically simulated power in the various forward-propagating bands as a function of pressure. In both graphs the signals are normalized to the input pump power. The purple dashed line in (b) shows the calculated AS2 power when the pump is initially launched in a perfect $LP_{01}$ mode and all intermodal coupling is switched off.

The propagation constants $\beta_{mn}$ for the different core modes of the kagomé-PCF, with $(m,n)$ being the azimuthal and radial mode orders, can be approximated by [20]:

$$\beta_{mn} = k_0\sqrt{n_{gas}^2(p,\lambda) - \lambda^2 u_{mn}^2/(2\pi a)^2} \quad (3)$$

where $\lambda$ is the wavelength, $k_0$ the vacuum wavevector, $n_{gas}$ the refractive index of the filling gas, $p$ is the gas pressure, $u_{mn}$ is the $n^{th}$ root of $m^{th}$ order Bessel function of the first kind and $a$ is the core radius.

Table 1. Maximum conversion efficiencies (CE) of the individual comb lines and their corresponding pressures.

| Sidebands | AS3 | AS2 | AS1 | S1 | S2 | S3 |
|---|---|---|---|---|---|---|
| CE (%) | 0.015 | 0.03 | 1.05 | 10.6 | 6.0 | 1.0 |
| P (bar) | 22 | 25 | 8 | 10 | 18 | 15 |

As already pointed out, we observed the higher-order Stokes and anti-Stokes comb-lines to emerge in a superposition of HOMs (see Fig. 3(a)). Among the various HOM contributions, the $LP_{11}$-like mode was the strongest, probably because it was hard to avoid exciting it with the pump light. For simplicity, therefore, we only include $LP_{01}$ and $LP_{11}$ modes in the model and disregard the losses of the comb lines, which are not significant for the fiber length used in the experiment (~10 cm). We set a noise floor of ~0.01 V/m for the different signals to obtain close agreement with the experiments; note however that different noise-floor levels do not affect the overall dynamics. With these considerations, we find the best overall agreement between the experimental results (Fig. 4(a)) and the simulations (Fig. 4(b)) for an $LP_{01}/LP_{11}$ ratio of ~4:1.

Nevertheless, there is a significant discrepancy between theory and experiment in the strength of the AS3 signal at 199 nm, the simulated AS3 signal being much weaker than the measured one, which has a complex near-field profile (see Fig. 3(a)) featuring high $LP_{02}$ mode content (not included in the simplified model). In general, we observe that the influence of the $LP_{02}$ and other HOMs increases for larger sideband orders, the $LP_{11}$ being dominant for sidebands closer to the pump. The largest difference between modeling and experiments, however, is in the total power and the S1 and S2 signals at high pressure (within the gray shaded area in Fig. 4(a)). In the modeling the S1 and S2 signals continuously increase with pressure, whereas in the experiment they saturate (S2 decays at higher pressures). Compared to the simulations, the experimental curve (solid-pink) for the total power contained in all bands significantly decreases (~3 dB) with increasing pressure. The reason is the onset of a strong backward-travelling Stokes signal that appears for pressures above 20 bar. This signal, which at high enough pressure was even stronger than the forward S1 signal, explains the measured decay of the total power in the forward direction, as well as the saturation of the S1 and S2 signals. As the model only describes the dynamics of forward propagating waves, it obviously overestimates the strength of the forward signals at high pressures since pump depletion by stimulated Raman backscattering is not included.

As we anticipated above, the presence of HOMs facilitates phase-matched ultraviolet light generation via molecular modulation. This relies on the fact that the strong $H_2$-gas dispersion at high pressures dominates over the waveguide dispersion for wavelengths well within the DUV/VUV range. This can be easily understood from Eq. 3, which shows that the gas dispersion increases with both pressure and frequency, while the waveguide dispersion (the second term under the square-root) decreases. Thus, $C_W$'s with large wavevectors are required for efficient generation of higher-order anti-Stokes sidebands. This is clear in the modal dispersion diagrams in Fig. 5. For a pressure of 25 bar, operation is in the normal dispersion regime, which precludes the possibility of all-$LP_{01}$ molecular modulation [17].

Interestingly, the dispersion profiles depicted in Fig. 5 suggest that the long wavevectors required for anti-Stokes generation might be provided by intermodal $C_W$'s involving Stokes signals in higher-order modes. A clear signature of such intermodal phase-matching is the double-humped structure of the AS2 signal observed in Fig. 4. The first AS2 peak occurs at roughly the same pressure as the AS1 peak (~10 bar), indicating cascaded up-conversion from a large reservoir of AS1 photons, which forces the two bands to follow similar dynamics with increasing pressure. The second broader peak at ~25 bar, however, is more interesting as it occurs when the strength of AS1 is already decreasing, suggesting that parametric amplification of the AS2 signal proceeds by a different phase-matching route than the first peak. We attribute this to the presence of an intermodal $C_W$ with wavevector $\beta_{01,11}$, created by the S1 to S2 conversion (solid green arrow in Fig. 5) which at ~25 bar phase-matches the AS1/$LP_{01}$ to AS2/$LP_{11}$ transition (dashed-green arrow in Fig. 5). This is corroborated by the near-field images in Fig. 3(a). Remarkably, the numerical simulations accurately reproduce this behavior (see Fig. 4(b)), revealing that the broadening and peaking of the signal in the pressure scan is due to the presence of some initial $LP_{11}$ pump content. This was confirmed by running a simulation where the pump is launched in a pure $LP_{01}$ mode and all nonlinear intermodal coupling is switched off. Under these conditions the second AS2 peak vanishes completely (purple dashed-line in Fig. 4(b)).

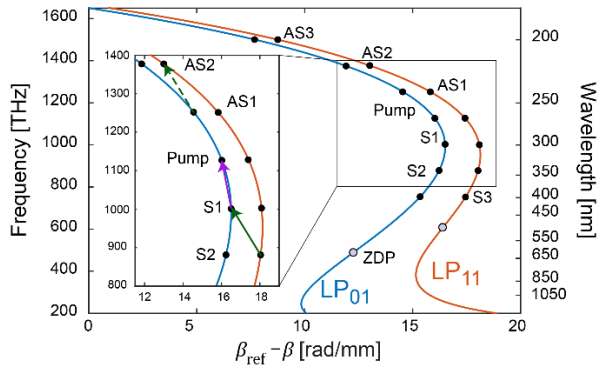

Fig. 5. Dispersion curves for the $LP_{01}$ and $LP_{11}$ modes in the experimental kagomé-PCF filled with 25 bar of $H_2$. The black dots mark the positions of the sidebands and the light gray circles indicate the zero dispersion points (ZDP) for each mode. The color-coded solid/dashed arrows represent the $C_W$'s (see text). To magnify the subtle features of the dispersion, we plot frequency against $\beta_{ref} - \beta$, where $\beta_{ref}$ is a linear function of pressure chosen such that $\beta_{ref} - \beta$ is zero at a frequency of 1650 THz.

Note that since the waveguide dispersion is much weaker than the gas dispersion at high frequencies, the propagation constants of different modes become more equal, enabling a given intermodal $C_W$ to excite mixtures of HOMs by molecular modulation. This explains the complex near-field mode profiles of the DUV/VUV sidebands.

We have suggested that the high-frequency extension of the comb is intimately linked to the appearance of higher-order Stokes components in HOMs. This is mainly triggered by the generation of intramodal $C_W$'s with wavevectors (see the $C_W$ with $\beta_{01,01}(\omega_P)$, represented by the solid pink arrow in Fig. 5) shorter than their intermodal counterparts. Nevertheless, once the higher-order Stokes bands are populated, different intra/intermodal $C_W$'s may contribute to the process.

In conclusion, a Raman comb of narrow bands from the VUV to the visible can be generated by pumping a $H_2$-filled kagomé-PCF at 266 nm and carefully adjusting the pressure. The system combines the unique guiding properties of kagomé-PCF in the DUV/VUV with the large Raman frequency shift and gain of $H_2$, enabling VUV wavelengths to be generated by a combination of SRS and intermodal molecular modulation at moderate pressures. The performance of this compact and potentially tunable fiber-based DUV/VUV light source could be further improved by using a spatial light modulator to accurately control the spatial profile of the pump beam, or by seeding the SRS process.